# Isostructural Phase Transition of TiN Under High Pressure

J.G. Zhao, L. X. Yang, Y. Yu, S. J. You, R. C. Yu, L. C. Chen, F. Y. Li, C. Q. Jin*

Institute of Physics, Chinese Academy of Sciences

X. D. Li, Y. C. Li, J. Liu

Institute of High Energy Physics, Chinese Academy of Sciences

Corresponding author: cqjin@aphy.iphy.ac.cn

**Abstract**

*In situ high-pressure energy dispersive x-ray diffraction experiments on polycrystalline powder TiN with NaCl-type structure have been conducted with the pressure up to 30.1 GPa by using the diamond anvil cell instrument with synchrotron radiation at room tempearture. The experimental results suggested that an isostructural phase transition might exist at about 7 GPa as revealed by the discontinuity of $V/V_0$ with pressure.*

Recently, titanium nitride has aroused intensive interests because of its simple structure and special properties such as high melting point, high hardness, high corrosion resistance, high specific strength, and metallic conductivity. Like most of refractory transition metal nitrides with NaCl-type structure, TiN is a superconductor with transition temperature about 5 K. $TiN_x$ keeps NaCl-type structure at a variant of N content (0.6<x<1.2), with different lattice constants and physical properties.[1] TiN has been mainly applied in hard coatings,[2] due to it exceptional combination of physical properties and chemical and metallurgical stability.[3] Some calculations have given detailed information about the electronic energy-band structure, indicating it is a conductor at ambient condition.[4-5] There are many experiments devoted to various aspects of titanium nitride film growth,[6-8] and the films could improve materials' properties and duration.[9] Although there exist many research efforts in films, the bulk properties of TiN under high pressure are less studied. The research for TiN with high-pressure method is essential to understand its high hardness and large bulk modulus. Here we report the investigation of high-pressure structural stability of TiN based on diamond anvil cell (DAC) technique by using *in situ* high-pressure energy dispersive x-ray diffraction with synchrotron radiation.

The *in situ* high-pressure x-ray energy dispersive diffraction experiment on TiN was carried out at room temperature in a diamond anvil cell at Beijing Synchrotron Radiation Facility (BSRF). The experiments details are similar to those reported [10-12]. The culet of diamond is 500 $\mu$m in diameter and the hole in a T301 stainless steel gasket is 300 $\mu$m in diameter. The powder sample is loaded into the hole in gasket and the platinum powder is covered with sample as an inner pressure standard. Here we used the equation of state of Pt to calculate the inner pressure of the DAC. The spot size of the focused x-ray beam was 120×120 $\mu$m and the storage ring operated at 2.2 GeV and 50-110 mA. In this experiment, the relation of energy of photon and channel was $E=0.50566+0.00839\times chn$. According to the (111) and (200) diffraction peaks of Pt and the formula $E(keV)\times d(Å)=6.19925/sin\theta$, we got the pressure and d spacing of sample at $2\theta=17.0°$.

Figure 1 shows the spectra of x-ray energy dispersive diffraction results of TiN under various pressures, the diffraction patterns are similar at every pressure. The diffraction data are collected in the pressure range of 0-30.1 GPa. There are three peaks of Pt: (111), (200) and (220) and three peaks of sample: (111), (200) and (220) at ambient pressure in x-ray energy dispersive diffraction pattern. There also are two escaping peaks indicated with signal * in low-energy region that result from the high rank of the peak (111) of Pt and the peak (200) of sample. All the Gaussian-type diffraction peaks shift with increasing pressure and return to original sites when the pressure releases to ambient. From the XRD pattern, the structural phase transition could not be found in this experimental pressure range because no splitting or merging peaks are observed. So the peak fitting, indexing and cell parameter refining are based on the NaCl-type structure of TiN with space group Fm-3m.

The variation of d values of peak (111), (200) and (220) of TiN with pressure is shown in Figure 2. The d value of peak (111) decreases with pressure below ~7 GPa, but increasing with pressure above 7 GPa. When the pressure exceeds ~11 GPa, it begins to decrease again. The d values of other peaks are decreasing all along with increasing pressure. The plot of volume compressibility vs. pressure shows a discontinuous change at around 7 GPa and there is a flat part in about 7-11 GPa, as shown in Figure 3. It is clear that there exists a big difference in slope and error bar of $V/V_0$-P curve between low pressure (below ~7 GPa) and high pressure (above ~11 GPa). Because no phase transition was found from XRD patterns, we thought that an isostructural phase transition exists at about 7 GPa.

It should be pointed out that in this reported phase transition induced by pressure, the structure type does not change, while the lattice constant or volume has a discontinuity with pressure resulting from the electronic topological transition, as those of isostructure phase transition observed in osmium[13], $K_8Si_{46}$ clathrate[14], $\alpha$- and $\beta$-$Ge_3N_4$[15], EuO[16], and so on. In the case of osmium, the changes in the band structure on compression could induce the changes of crystal structure and the maxima in band $h_7$ below the Fermi energy will pass through the Fermi surface resulting in the collapse of small hole ellipsoid under high pressure, which contributes to the isostructural phase transition.[13] The *ab initio* phonon band structure calculations have revealed the phonon instabilities on $K_8Si_{46}$ and other materials while the phase transition happened under high pressure.[17,18] The EuO with NaCl-type structure undergone an isostructural phase transition at near 30 GPa and became CsCl-type structure at about 40 GPa.[16] When the first phase transition happened, EuO kept the primary crystal symmetry but the volume dropped by about 4%, which was conducted by the electronic collapse in the Eu ion involving the 4f-5d electron transition. There is also other sort of isostructural phase transition induced by the change of temperature in GdPdAl single crystals reported by Kusz et al.[19] The authors thought that the different movements of each atom layer under low temperature could induce the changes of lattice constants. Although there is no conclusive explanation to the change, the isostructural phase transition is believed a novel phenomenon in some materials as a function of variables of pressure and temperature.

The structural phase transition of materials in NaCl-type has been studied in thermal dynamics.[20] There were several types of new structure, such as rhombohedra (space group: R-3m), CsCl-type (space group: Pm-3m). It was found that some NaCl-type LnX (Ln is lanthanide and X=P, As, Sb, Bi) would become normal or distorted (tetragonal) CsCl-type structure under high pressure through *in situ* high-pressure powder x-ray diffraction experiments[21,22,23] Two phases existed around transition pressure in LnX materials, which indicated the phase transition was a gradual process. We speculate it is possible for TiN to have phase transition under high-pressure analogue to the above-mentioned case. Under higher pressure, TiN may become CsCl-type structure or other phases like EuO.[16] But we could not get further more about the phase transition on TiN from our limited XRD patterns. The detailed structure refinement is needed based on more accurate high-pressure angle resolved x-ray or neutron experiments.

**Acknowledgements:**
This work was supported by the National Science Foundation, and the Ministry of Science and Technology of China (2002CB613301).

**Figure captions:**

**Fig. 1.** The spectra of energy dispersive X-ray diffraction pattern of TiN at room temperature from ambient up to 30.1 GPa.

**Fig. 2.** The pressure dependence of *d* evolution for TiN.

**Fig. 3.** Volume compressibility versus pressure of TiN.

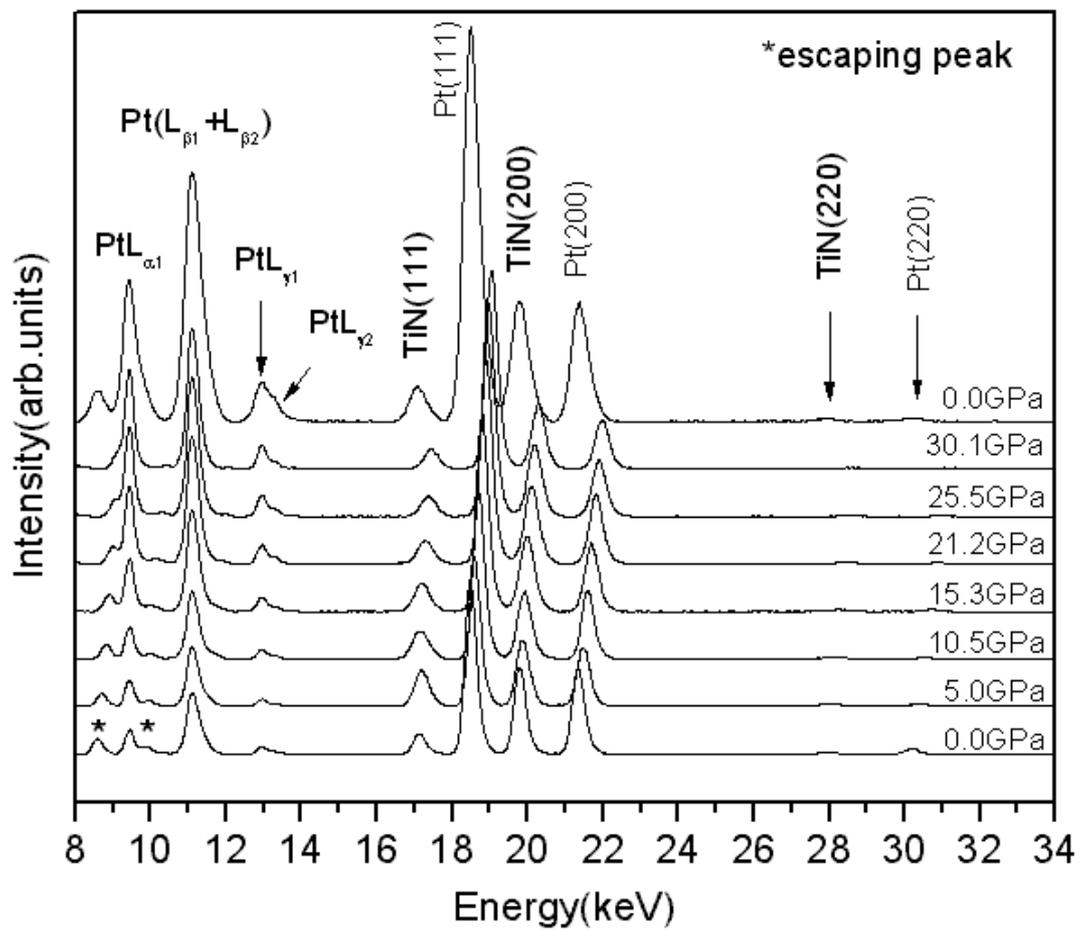

**Fig. 1.**

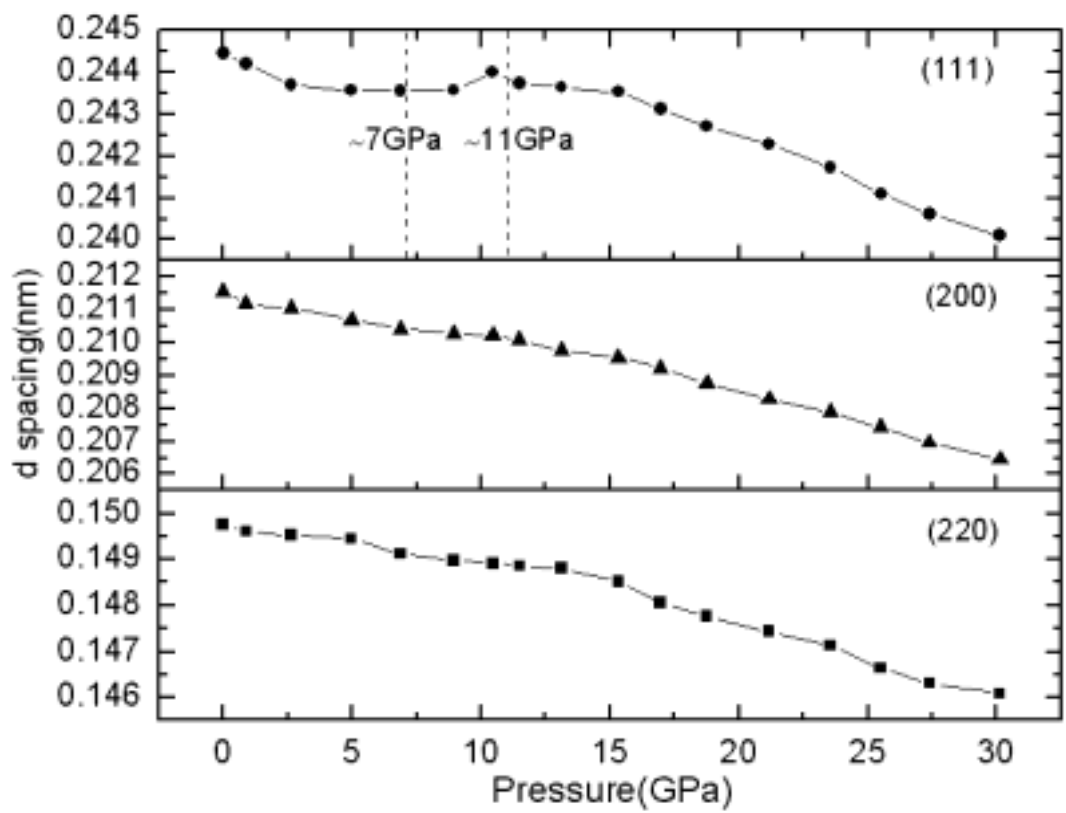

**Fig. 2.**

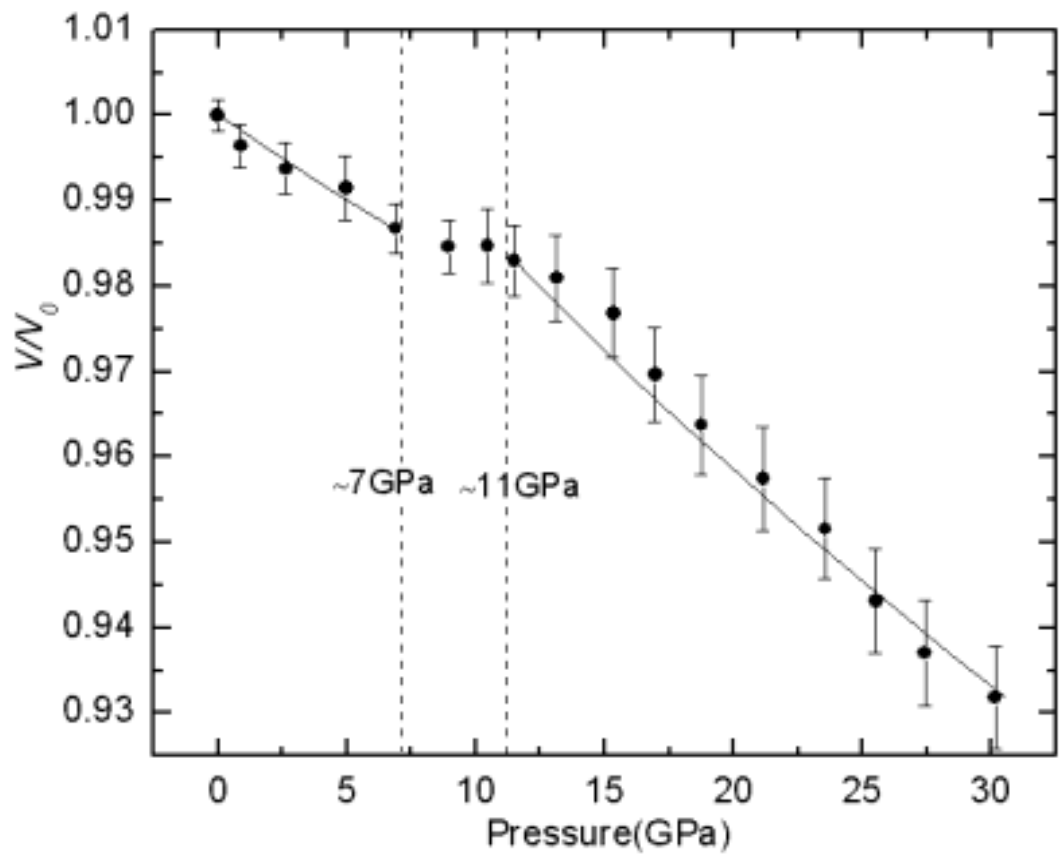

**Fig. 3.**